*Original Article*

# Bridging Climate Awareness and Sustainable Entrepreneurship: A Conceptual Framework Based on the Theory of Planned Behavior

[1]**Muhammad Rofiqul Islam**, [2]**Abdullah Al Mehdi**
[1]*A.R. Sanchez, Jr. School of Business, Texas A&M International University, Texas. USA.*
[2]*Starco Impex, Inc., Texas, USA.*

***Abstract:*** *Many studies have examined the connection between the intention to start a business and environmental values. However, there still needs to be more knowledge in the extant literature about how climate change campaigns influence sustainable entrepreneurial intention. This study uses the Theory of Planned Behavior (TPB) to develop a theoretical framework to explain how climate change campaigns affect the intention to start a sustainable business. This interdisciplinary conceptual research model bridges the gap between climate awareness, sustainable values, and entrepreneurial intentions, offering a robust framework for understanding and fostering sustainable entrepreneurial behaviors. Our study lays the groundwork for future empirical studies and real-world interventions to advance sustainability through entrepreneurship.*

## I. INTRODUCTION

Living standards and community welfare can be improved by engaging in entrepreneurial activities, as those activities drive economic growth (Shane & Venkataraman, 2000). The economic growth concept has gained new momentum with increased attention on the new concept of sustainable development, which ensures present economic growth without compromising the growth potential for the future generation to meet their needs (Brundtland, 1987; Hernández-Perlines et al., 2017). Many recent academic studies have investigated the connection among sustainable growth and entrepreneurship, emphasizing entrepreneurship's innovative capacity to usher in the next industrial revolution and a new sustainable future, where entrepreneurship has taken on new dimensions such as green entrepreneurship, social entrepreneurship, environmental entrepreneurship, ecopreneurship, and sustainable entrepreneurship. The rise of driven by meaning, for-profit companies has been evident in the past few decades to demonstrate entrepreneurial answers to the environmental and socioeconomic problems. This is reflected in the origination and development of a vibrant research subdomain termed "sustainable entrepreneurship." The founding principle of sustainable entrepreneurship is that preserving the ecological and social environment must be a priority in undertaking economic activities and restoring the environmental balance among nature, society, and economic activities. The commitment of sustainable entrepreneurship is to integrate the economic activities in society in such a manner that ensures sustainable development (Anderson, 2000; Berle, 1993; Blue, 1990; Dean & McMullen, 2007; Dees, 1998; Keogh & Polonsky, 1998; Lans et al., 2014; Parrish, 2010; Shepherd & Patzelt, 2011; Stubbs, 2017). The ability of a sustainable entrepreneurial approach to combine economic, social, and environmental values in the creation process is necessary for future generations' well-being (Hockerts & Wüstenhagen, 2010).

Individual behavioral patterns are necessary to be an entrepreneur and are shaped by entrepreneurial purpose, a driving factor. Considering the importance of entrepreneurial intention, research attention is focused mainly on the study of entrepreneurial intention. Entrepreneurship researchers are interested in examining the aspects that lead to individual entrepreneurial orientation since individuals interact with their environments to acquire these entrepreneurial intentions (Liñán et al., 2011; Shi et al., 2020; Zhao et al., 2010). It is necessary to explore the antecedents of individual entrepreneurial approaches by looking at how personal values influence entrepreneurial conduct.

Schwartz (2011) claims that social adaptation and survival are contingent on fulfilling specific biological and psychological needs influenced by individual values. Thus, the expected relationship between individual values and entrepreneurial intentions is supported by rational logic. Empirical studies also showed a link between an individual's values and how they act as an entrepreneur, but Schwartz's (2012) theory of individual values is rarely used in the literature on entrepreneurship. Some researchers use certain types of individual values to explain individual entrepreneurial intentions.

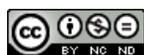

Noseleit (2010) showed the differences in values between entrepreneurs and non-entrepreneurs. A more comprehensive analysis of Schwartz's (2012) theory of individual values in shaping entrepreneurial orientation needs to be conducted. Individual values are used to explore the effect of individual values on individuals' entrepreneurial attitudes and social norms (Morales et al., 2019; Schwartz et al., 2012b) and how entrepreneurial attitudes and social norms contribute to individual entrepreneurial orientation (IEO).

Individuals are influenced by their personal values to participate in sustainable entrepreneurial endeavors. An entrepreneur's environmental ideals are a necessary precondition for their sustainable entrepreneurial goal (Kaesehage et al., 2019; Peng et al., 2021). The study aims to explore the effects of the climate change campaign on shaping environmental values and the ultimate effects on sustainable entrepreneurial intentions by influencing individual environmental values.

## II. LITERATURE REVIEW

### A) Sustainable Development Values

Sustainable development values are a combination of fundamental human values that share and influence certain attitudes and behaviors toward achieving sustainable development goals in the twenty-first century. In the twenty-first century, these principles of sustainable development shape the dynamics of global interactions and provide governments and organizations to design their policies in conformity with sustainable development goals. In addition, these sustainable development values influence the behavioral patterns of organizations and individuals, ultimately contributing to the quality of people's lives and the environment (Assembly, 2000; Shepherd et al., 2009; Shepherd & Patzelt, 2011). From what we have discussed, the main goal of evaluating sustainable development values is to improve people's lives and protect the environment.

*Freedom:* People, irrespective of gender, can choose their own lives and raise their children with human dignity. They have the right to live hunger-free and violence-free lives. These values also advocate for participatory democracy to achieve freedom in our lives.

*Equality:* Everyone, no matter who they are, where they come from, or their gender, should have the same access to development benefits.

*Solidarity:* Fairness and justice should come first when figuring out how to share the costs and benefits of dealing with global problems.

*Tolerance:* People should respect and value the fact that people have different beliefs, cultures, and languages.

*Respect for nature:* This value shows how important it is to take care of nature by managing natural resources well to ensure sustainable development.

*Shared responsibility:* These values represent a multilateral economic and social development approach.

The above-noted fundamental values related to sustainable development are consistent with the two production of social and environmental value is one facet of sustainable enterprise. Some of the sustainable development values are related to the creation of social values, such as tolerance, solidarity, freedom, and equity. In contrast, others are related to creating environmental values, such as respect for nature and shared responsibility. We expect a link between sustainable development values and the intention to be a sustainable entrepreneur since their goals for behavior are similar.

### B) Climate Change Campaigns and Sustainable Development Values

Climate change awareness among stakeholders is regarded as one of the most critical keys to addressing the challenges posed by climate change (Tan et al., 2008). Climate change campaigns are meant to make people more aware of the issue, change how they respond to it, and change how they act.

People's sustainability values drive their attitudes and behaviors toward sustainable development. The UN declaration identifies specific values about sustainable development, such as freedom, equality, solidarity, tolerance, respect for nature, and shared responsibilities. Understanding the nature and ecological functions of human behaviors allows us to interpret people's strategic actions, attitudes, and behaviors about sustainable development efforts (Assembly, 2000; Loomes, 2006; Shaikh et al., 2007; Spash, 2002; Thøgersen & Ölander, 2002). Human values must be refined and redesigned to form a new set of sustainable development values. These help individuals redirect and redesign their goals and attitudes toward sustainable development (Leiserowitz et al., 2006; Saifi & Drake, 2008). Climate change campaigns can contribute to creating sustainable development values with enhanced climate awareness, which ultimately helps change people's attitudes and behaviors. As a climate change campaign aims to increase public awareness about climate change's effects, people are expected to demonstrate sustainable development values and share their attitudes and behaviors accordingly. Climate change awareness drives individuals and managers to change their behavioral patterns and subsequent actions (Halady & Rao, 2010; Raducu et al., 2020). Hence, our first proposition is:

*Proposition 1: A campaign about climate change has a positive effect on sustainable development values by making people more aware of climate change.*

### C) Climate Awareness and Behavioral Orientation

Research shows that approximately half of the American public is not convinced about the adverse effects of climate change and, therefore, is not pursuing behavior change toward mitigating climate change impacts through changing behavioral orientation. Empirical evidence from the UK data also shows that lack of knowledge, skepticism, and distrust in information sources influence people's behavior toward climate change mitigation efforts (Lorenzoni et al., 2007; Maibach et al., 2009).

It is okay to conceptualize and express climate change issues by creating messages that are meant to incite dread. On the other hand, a proven climate change campaign illustrates the gravity of the situation and the threat it poses, and it has been shown to be a successful program that can motivate people to take action. In addition to mentioning the threats and actions, the message should communicate the potential benefits of the intended action, and the role of the individual should also be mentioned (Nisbet, 2009; Patchen, 2006). We discovered from the preceding discussion that climate awareness influences individuals' behavior in combating the effects of climate change. When entrepreneurs learn about climate change, they can change their actions to be more sustainable.

### D) Sustainable Entrepreneurship

A new perspective on entrepreneurship views entrepreneurial activities as a mechanism to manage and restore the environment rather than blaming business activities for environmental damage and social inequality. This new perspective encourages us to identify and appreciate sustainable enterprise as a novel form of entrepreneurship (Muñoz & Cohen, 2018).

The central component of sustainable entrepreneurship is the individual sustainable entrepreneur and their behavioral patterns associated with their intention to pursue sustainable entrepreneurial initiatives. Academic studies also focus on exploring and defining factors that affect individuals' behavior and encouraging them to initiate sustainable entrepreneurship. As a result, this entrepreneurial intent is considered the key to determining the likelihood of starting an entrepreneurial venture. The individual's desire to be a sustainable entrepreneur and corresponding attitudes contribute to sustainability-oriented entrepreneurial intent (Van Oorschot et al., 2018). This entrepreneurial intent inspired by alternative motives and values may substantially influence the socioeconomic landscape toward sustainable development (Tilley & Young, 2009).

### E) Environmental Values and Sustainable Entrepreneurial Intention

Entrepreneurs demonstrate their entrepreneurial characteristics by deliberate action, not just having a particular state of mind; actions must be pursued. These entrepreneurial actions reflect behavioral patterns originating from or guided by entrepreneurial intention. Entrepreneurial intention is a deliberate state of mind that defines and demonstrates certain behaviors, priorities, and emotions toward planned entrepreneurial actions. Thus, entrepreneurial intention is regarded as the most critical predictor of entrepreneurial action, and we can reasonably forecast entrepreneurial actions based on the presence of entrepreneurial intention (Bird & Jelinek, 1989; Krueger Jr et al., 2000; Maaloui et al., 2018; McMullen & Shepherd, 2006).

Individuals' values related to sustainable development can predict attitudes, perceptions, and behavioral patterns toward sustainable entrepreneurship. Individuals' environmental values not only influence their intention to encourage their entrepreneurial activities to have a positive environmental impact while simultaneously fostering a rising consciousness, which leads to the development of new core values about sustainable entrepreneurship (Nuringsih et al., 2019; Nuringsih & Puspitowati, 2017). The environmental core values drive entrepreneurs and potential entrepreneurs to boost their intention to create economic value by ensuring environmental value creation (Kuckertz et al., 2019).

Companies are significant contributors to environmental degradation and the victims of environmental deterioration as climate change increases resource constraints regarding availability and costs. Entrepreneurs are being directed toward sustainable entrepreneurship to deal with environmental challenges. Individuals interested in new ventures also shape their entrepreneurial intentions by considering economic, social, and environmental value creation. The behavioral orientation of a sustainable entrepreneur is influenced by the motivations to become a sustainable entrepreneur (Arru, 2020; Mbebeb, 2012; Peng et al., 2021; Sher et al., 2020). Considering the importance of environmental values in shaping an entrepreneur's intention to explore nature, academics need to focus on the extent to which environmental principles and sustainable entrepreneurial goals are related.

The integration of sustainability issues with their entrepreneurial activities is affected by individuals' environmental values, which are considered the major driving forces toward sustainable entrepreneurship. However, integrating sustainability into the business processes takes work, as economic value creation conflicts with environmental value creation, and lower economic profit potential may pressure an entrepreneur's sustainable entrepreneurial intentions. Individuals' environmental values help to understand attitudes toward sustainable entrepreneurship. It creates two mutually reinforcing processes:

environmental value creation and adopting sustainable entrepreneurship. Considering these processes, research attention should be directed toward exploring the determinants of these processes. Climate change campaigns can contribute to inducing the environmental value creation process, whereas the Theory of Planned Behavior (TPB) can be used to explain sustainable entrepreneurial intentions.

*Proposition $2_a$: Sustainable development values are positively associated with entrepreneurial attitudes.*
*Proposition $2_b$: Sustainable development values are positively associated with entrepreneurial subjective norms.*
*Proposition $2_c$: Sustainable development values are positively associated with perceived behavioral control.*

The TPB theory is superior to other theories in explaining behavioral intention, so its use to predict sustainable entrepreneurial behavior is justified. According to the TPB, three direct predictors (attitude, subjective norm, and perceived behavioral control) of behavioral intention can be measured by a set of predefined and validated statements. The validity of the theory developed using the TPB can be empirically tested. In addition to this, the models developed using the TPB can efficiently predict behavioral intention in different contextual settings. With the TPB theory, we can easily use the theory to predict sustainable entrepreneurial intention in terms of behavioral predictors.

Moreover, the TPB theory is flexible in structure, so researchers can easily incorporate new predictors or mediators into the model (Ajzen, 2002; Riebl et al., 2015; Schwartz, 1977; Timm & Deal, 2016). The study's research design requires some level of flexibility because we are trying to develop a new model by introducing new variables and moderators (climate change campaign, climate awareness). Our approach of introducing new variables in the TPB model to explain sustainable entrepreneurial behavior is inspired by the success of previous studies. In one such meta-analysis (Yuriev et al., 2020), the author found that about two-thirds of the research, articles used a modified TPB model by introducing a new variable as a direct predictor of behavioral intention. We propose a model (Figure 1) for explaining sustainable entrepreneurial behavior by extending the TPB theory with a set of antecedents (individual environmental values) and mediating variables (climate change campaigns and climate awareness).

*Proposition $3_a$: Climate awareness moderates the relationship between sustainable development values and attitudes towards sustainable entrepreneurship.*
*Proposition $3_b$: Climate awareness moderates the relationship between sustainable development values and subjective norms.*
*Proposition $3_c$: Climate awareness moderates the relationship between sustainable development values and perceived behavioral control.*

### F) Theory of Planned Behavior and Sustainable Entrepreneurial Intention

According to the TPB model, any planned behavior is the result of some type or form of intentionality toward that behavior. Intentionality reflects individuals' motivational factors, where they demonstrate different levels of willingness to engage in a specific behavior. A higher level of intention means a higher motivation to engage in relevant behavioral demonstrations. Here, we consider sustainable entrepreneurial intentions as specific intentions to undertake sustainable entrepreneurial initiatives, and the TPB can easily explain this relationship. According to the theory, having entrepreneurial intention means being willing to do what you need to do to be a successful entrepreneur.

The TPB used extensively in academic research, can reasonably predict how entrepreneurs or people who want to be entrepreneurs will act and their long-term plans (Lortie & Castogiovanni, 2015). With the help of the TPB, we can explore the behavioral patterns and underlying individual values related to sustainable entrepreneurial intention. Research studies show that attitude demonstrates entrepreneurs' ability to evaluate their entrepreneurial behavior, and with this evaluation, they shape their behavior toward entrepreneurial intention. Entrepreneurial behavior is also contingent on the relational aspect of entrepreneurs, whose entrepreneurial behaviors are influenced by the pressures they feel from society and the environment (Kuckertz & Wagner, 2010; Lortie & Castogiovanni, 2015; Thelken & de Jong, 2020). Social and environmental pressures on individuals also shape entrepreneurial behavior, which can be explained under the "social norms" of the TPB (Mbebeb, 2012). Social norms reflect the entrepreneurial behavior that results from perceived social pressure (Ajzen, 1991).

Among the conceptual frameworks for studying human action, the theory of planned behavior (TPB) has emerged as one of the significantly accepted frameworks in behavioral studies due to its ability to explain behavior in terms of beliefs about the behaviors (Ajzen, 1991, 2001, 2005). According to the TPB, three major categories of beliefs about an action guide an individual's decision to be oriented towards the action or not: behavioral beliefs, normative beliefs, and control beliefs. First, their belief categories include behavioral beliefs, which demonstrate an individual's beliefs about the potential consequences of the action under consideration. When individuals believe that the performance of an action will bring favorable consequences for them, they are likely to be oriented toward action. On the other hand, a negative belief about the consequence is likely to result in withdrawal from the action. It may be viewed as a behavioral evaluation of the potential action's perceived consequences.

These behavioral beliefs, formed by evaluating the consequences of an action, constitute a favorable or unfavorable attitude toward the behavior. Second, individuals' behavioral orientation is also influenced by their normative beliefs about other people's expectations. According to the integrated social contract theory (ISCT), common beliefs or consensuses influence individuals' behavioral orientation. So, both theories support the idea that the expectations of other close relatives influence individuals' behavioral orientation about an action. In the TPB, these kinds of beliefs are referred to as "subjective norms." Third, people consider the factors that may help or hinder the performance of an action. The levels of control over the influencing factors contribute to different behavioral consequences. When people believe they have greater control over the influencing factors, their behavioral orientation is expected to be toward carrying out that action. These control beliefs represent perceived behavioral control, influencing the behavioral orientation towards action.

*Attitude:* The direct link between attitudes and behaviors is not established in academic research; however, behaviors can reasonably be predicted by attitudes and beliefs through behavioral intention, where attitudes and beliefs have a mediating relationship with behaviors (Kolvereid, 1996). Individuals' attitudes and beliefs related to sustainable entrepreneurial behaviors through entrepreneurial intention

The term "attitude" one holds towards a behavior can be described as how an individual views a behavior in terms of his evaluation. A positive evaluation of the behavior may also be termed a positive attitude towards that behavior, which ultimately constitutes a favorable intention to demonstrate behavior for which they have a favorable intention (Ajzen, 1991). Individuals must evaluate sustainable entrepreneurship positively to develop a positive intention toward it. When individuals consider being entrepreneurs for economic and environmental value creation a better choice, they form a sustainable entrepreneurial intention. The basic premise of attitude is derived from the expectancy-value model, which describes that the strength of beliefs is positively associated with how the subjective value affects the attitude toward a given outcome, such as entrepreneurial initiatives (Fishbein & Ajzen, 1977). When individuals have strong positive beliefs or evaluations about sustainable entrepreneurship, there is a higher probability that they will have sustainable entrepreneurial intentions. After figuring out our solid and positive entrepreneurial beliefs, we must investigate or evaluate what affects those beliefs.

*Subjective Norms:* The TPB supports the view that individuals' normative beliefs about a specific behavior influence behavioral orientation. The normative beliefs may be explained in terms of two types of normative beliefs: injunctive normative beliefs and descriptive normative beliefs (Fishbein & Ajzen, 2010). The former type represents the expectation or subjective probability that others approve or disapprove of the potential or actual performance of the behavior under consideration. The latter type describes beliefs that other people also perform the behavior under consideration. The injunctive and descriptive normative beliefs contribute to perceived pressure from close relationships and society.

*Perceived behavioral control:* Individuals' will or intentions can reasonably explain human social behavior. However, a situation may also exist where individuals have less control over behavioral orientation. Behavioral orientation is a situation that may not be explained by individuals' volitional control (Fishbein & Ajzen, 1977). Perceived behavioral control can be used to explain situations in which people have less volitional control over their behavior than they would like. Perceived behavioral control explains the level of control over the factors that influence the consequences of behavioral orientation. Ordinary everyday behavior is also subject to unforeseen environmental factors that limit volitional control over behavior. Perceived behavioral control measures the degree of control over the behavior, which affects the intention to demonstrate the behavior. A high level of perceived control is expected to increase a person's intention to perform the behavior. The above argument helps to construct the following propositions:

*Proposition 4: Attitudes, subjective norms, and perceived behavioral control are positively associated with sustainable entrepreneurial intention.*

Conceptual model

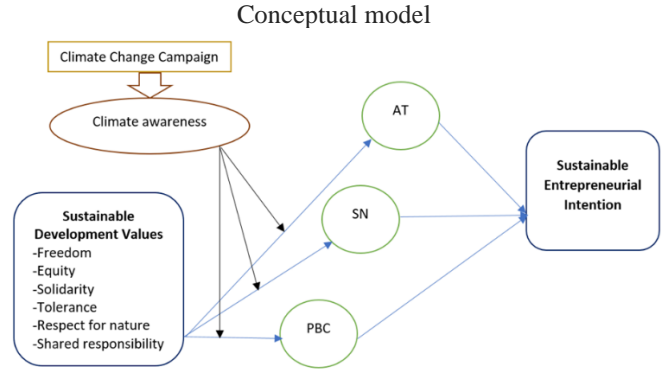

Note: AT: Attitude; SN: Subjective norms; and PBC: Perceived behavioral control
**Fig. 1 Conceptual Model**

## III. DISCUSSION

Our conceptual model delves into the complex interrelationships of campaigns against climate change, values for sustainable development, entrepreneurial mindsets, and aspirations for sustainable enterprise. The approach is organized around several hypotheses that clarify how awareness and values propel sustainable entrepreneurial actions. This study shows the importance of knowledge sharing in forming values consistent with sustainable development objectives. Campaigns against climate change can raise public awareness of the environmental crisis and promote a sustainable value system by utilizing various media platforms and educational initiatives. This is consistent with previous research showing how focused communication tactics can change environmental attitudes and behaviors (Moser, 2010).

People who internalize the principles of sustainable development are more likely to view supportive social norms, have positive entrepreneurial attitudes, and feel more in control of their entrepreneurial behavior. This aligns with the Theory of Planned Behavior, which holds that the main predictors of behavioral intentions are attitudes, subjective standards, and perceived behavioral control. The model extends this theory into sustainable entrepreneurship by incorporating sustainable development ideals into these constructs. This emphasizes the significance of contextual awareness in fortifying the relationship between values and entrepreneurial goals. By increasing the significance of sustainable development principles in entrepreneurial decision-making processes, climate awareness can strengthen the influence of these principles. This implies that initiatives to raise public knowledge of climate change may improve how sustainability-focused value education encourages entrepreneurial behavior. The intention to pursue sustainable entrepreneurial initiatives is significantly influenced by attitudes towards sustainable entrepreneurship, perceptions of social pressure to engage in such behavior, and self-assurance in one's capacity to carry out sustainable entrepreneurial activities. These connections highlight the importance of encouraging surroundings that foster optimistic outlooks, encourage long-term standards, and foster an entrepreneurial spirit.

The conceptual paradigm in this study makes critical theoretical advancements through incorporation. A thorough study of the causes of sustainable entrepreneurial aspirations is possible through the combination of recognized behavioral frameworks with values for sustainable development and awareness of climate change. Successful campaigns may play an essential part in raising a new generation of sustainable entrepreneurs against climate change and educational initiatives that support the principles of sustainable development. This conceptual study model provides a robust framework for comprehending and encouraging sustainable business behaviors by bridging the gaps between climate knowledge, sustainable ideals, and entrepreneurial goals. The ideas presented here lay the groundwork for upcoming empirical studies and real-world implementations that seek to advance sustainability via entrepreneurship.

## III. THEORETICAL AND PRACTICAL IMPLICATIONS

Our research makes critical theoretical contributions to sustainable development, entrepreneurship, and environmental psychology. This model expands on current theoretical frameworks by incorporating the Theory of Planned Behavior with values for sustainable development and knowledge of climate change, specifically in the context of sustainable entrepreneurship. A more sophisticated understanding of how environmental values and awareness translate into sustainable entrepreneurial intentions and behaviors is made possible by this interdisciplinary approach. First, the model emphasizes how crucial climate change campaigns are to raise public awareness and promote sustainable development principles. This supports and builds on an earlier study (Moser, 2010) highlighting the influence of focused communication tactics on environmental attitudes and behaviors. The concept emphasizes the potential for climate awareness to influence individual behaviors and stimulate more extensive economic operations that promote sustainability by placing these campaigns within the framework of entrepreneurial action.

Second, the model illustrates how firmly held values can impact entrepreneurial mindsets by connecting sustainable development principles with entrepreneurial views, arbitrary standards, and the impression of behavioral control. This extension of the Theory of Planned Behavior into sustainable enterprise suggests the incorporation of values into the development of entrepreneurial intentions. This theoretical development lays the groundwork for upcoming empirical studies investigating how values impact entrepreneurial behavior.

Third, by including climate awareness as a moderating factor, the impact of values on entrepreneurial ambitions might be amplified by contextual circumstances, which adds a new dimension to our understanding. This demonstrates the dynamic interaction between personal views and outside factors; raising knowledge of climate change could improve the congruence of ethical principles with entrepreneurial conduct.

This conceptual model has broad practical applications and can provide insightful information to practitioners, educators, and politicians who want to encourage sustainable entrepreneurship. According to the model, climate change campaigns can be helpful for policymakers in encouraging a sustainable culture. These efforts can instil principles that support sustainable development by increasing public awareness of climate challenges. This, in turn, may encourage more sustainable entrepreneurship. Therefore, to reach a broad audience, policymakers should invest in comprehensive communication strategies that use various media outlets and educational activities. Educators can use the approach to create curricula that combine entrepreneurship instruction with sustainable development principles. Teachers can raise a new generation of entrepreneurs motivated to solve environmental problems through creative business solutions and who are conscious of environmental issues by including sustainability ideas in entrepreneurship teaching. This strategy can support the development of a mindset that values both sustainability and enterprise, bridging the gap between environmental awareness and functional entrepreneurial abilities. The approach emphasizes for practitioners the significance of establishing supportive settings that perpetuate sustainable ideals, especially for those engaged in entrepreneurship support and growth. Offering tools, networking opportunities, and mentoring that highlight sustainability as a fundamental business principle are a few ways to do this. Programs that increase self-assurance in sustainable entrepreneurship can also improve behavioral control perception, which raises the possibility of sustainable businesses succeeding. This conceptual model's theoretical developments and real-world implementations offer a strong foundation for comprehending and encouraging sustainable entrepreneurial behaviors. The model creates the framework for upcoming empirical research and practical interventions that seek to advance sustainability through entrepreneurship by bridging the gaps between knowledge about climate change, sustainable ideals, and entrepreneurial aims.

### IV. FUTURE RESEARCH AVENUES

The paper presents a conceptual model that provides a thorough framework for comprehending the interrelationships among climate change campaigns, values related to sustainable development, and aspirations for sustainable entrepreneurship. It also opens several avenues for future research. First, the proposed conceptual relationships may be verified by empirical studies. It is possible to quantify the effect of climate change campaigns on sustainable development values and subsequent entrepreneurial intents by using quantitative research approaches such as surveys and experiments. Longitudinal studies would be particularly beneficial in monitoring how these associations evolve and finding potential causal ties. Second, examining the model in various cultural, economic, and demographic contexts would improve the ability to apply the findings to a broader range of situations. Comparative studies could investigate the impact of differences in climate awareness and sustainable development values on entrepreneurial intents across different locations and diverse population segments, including age groups, educational backgrounds, and socioeconomic situations. Third, additional investigation is required to reveal the fundamental mechanisms by which climate change campaigns impact sustainable development values and entrepreneurial intents. Qualitative methodologies, such as interviews and focus groups, have the potential to offer a more profound understanding of the cognitive and emotional mechanisms that underlie these behaviors. A comprehensive understanding of these mechanisms would provide a more detailed and precise direction for developing impactful campaigns and educational Programs. Fourth, exploring supplementary moderating and mediating factors could enhance our comprehension of the linkages within the model. For example, variables, including variations in environmental concern among individuals, previous experience in entrepreneurship, and resource availability, could influence sustainable development values' effect on entrepreneurial intents. It is essential to investigate mediating variables, such as information acquisition and perceived social support, to understand their impact on the relationship between values and behavior. Fifth, although this study utilizes the Theory of Planned Behavior, incorporating more theoretical views could enhance the model even more. Theories such as social cognitive theory, which highlights the importance of self-efficacy and observational learning, and the norm activation model, which centers around personal norms and altruistic behaviors, could offer further understanding of the factors that influence sustainable entrepreneurial actions.

Moreover, further investigation should also prioritize developing and evaluating targeted interventions derived from the model. Experimental studies can evaluate the efficacy of various climate change initiatives and educational Programs in promoting sustainable entrepreneurial aspirations. In addition, policy-oriented research should assess how governmental and institutional assistance can enhance the implementation of sustainable entrepreneurship practices.

## V. CONCLUSION

This study introduces a conceptual framework that explains the intricate relationship between climate change campaigns, values related to sustainable development, attitudes associated with entrepreneurship, and aspirations to engage in sustainable entrepreneurship. The model combines the Theory of Planned Behavior with sustainable development ideals and climate change awareness to create a complete framework for understanding how awareness and values influence sustainable entrepreneurship behaviors. The model's propositions emphasize the following key relationships: climate change campaigns have a positive influence on sustainable development values by increasing awareness; sustainable development values are connected to entrepreneurial attitudes, subjective norms, and perceived behavioral control; and climate awareness moderates these relationships, amplifying their effect. Moreover, attitudes, subjective norms, and perceived behavioral control influence sustainable entrepreneurial ambitions. This model makes substantial theoretical advances by expanding the Theory of Planned Behavior to include sustainable business. The model provides a more comprehensive understanding of the elements that impact sustainable entrepreneurship intents and behaviors by integrating climate awareness and sustainable development values. This interdisciplinary approach enhances the current body of literature by connecting the fields of environmental psychology, entrepreneurship, and sustainable development, therefore filling in the gaps between them. The model offers significant information for policymakers, educators, and practitioners. Policymakers can utilize climate change campaigns to promote a culture of sustainability and stimulate entrepreneurial endeavors that contribute to sustainable development. Educators can create educational plans that combine sustainable development principles with entrepreneurship instruction, thereby fostering a new cohort of entrepreneurs that prioritize sustainability. Practitioners can establish environments that promote sustainable principles, offer resources, guidance, and networking opportunities, and instil confidence in sustainable business endeavors.

Overall, this conceptual model provides a strong basis for future empirical investigations and practical interventions that seek to foster sustainability through entrepreneurship. The model provides a complete framework for understanding and promoting sustainable entrepreneurial behaviors by connecting climate knowledge, sustainable principles, and entrepreneurial objectives. This paradigm not only enhances theoretical understanding but also offers practical tactics for promoting sustainable entrepreneurial initiatives, ultimately contributing to the overarching objective of sustainable development.

## VI. REFERENCES

[1] Ajzen, I. (1991). The theory of planned behavior. *Organizational Behavior and Human Decision Processes*, *50*(2), 179–211.
[2] Ajzen, I. (2001). Nature and operation of attitudes. *Annual Review of Psychology*, *52*(1), 27–58.
[3] Ajzen, I. (2002). Perceived behavioral control, self-efficacy, locus of control, and the theory of planned behavior 1. *Journal of Applied Social Psychology*, *32*(4), 665–683.
[4] Ajzen, I. (2005). *EBOOK: Attitudes, Personality and Behaviour*. McGraw-hill education (UK).
[5] Anderson, T. L. (2000). *Enviro-capitalists: Doing good while doing well*. Rowman & Littlefield Publishers.
[6] Arru, B. (2020). An integrative model for understanding the sustainable entrepreneurs' behavioural intentions: An empirical study of the Italian context. *Environment, Development and Sustainability*, *22*(4), 3519–3576.
[7] Assembly, U. G. (2000). United Nations Millennium Declaration. *United Nations General Assembly*, *156*.
[8] Berle, G. (1993). *The green entrepreneur: Business opportunities that can save the Earth make you money*.
[9] Bird, B., & Jelinek, M. (1989). The operation of entrepreneurial intentions. *Entrepreneurship Theory and Practice*, *13*(2), 21–30.
[10] Blue, J. (1990). *Ecopreneuring: Managing For Results (London: Scott Foresman)*.
[11] Brundtland, G. H. (1987). Our common future—Call for action. *Environmental Conservation*, *14*(4), 291–294.
[12] Dean, T. J., & McMullen, J. S. (2007). Toward a theory of sustainable entrepreneurship: Reducing environmental degradation through entrepreneurial action. *Journal of Business Venturing*, *22*(1), 50–76.
[13] Dees, J. G. (1998). *The meaning of social entrepreneurship*. Kauffman Center for Entrepreneurial Leadership.
[14] Fishbein, M., & Ajzen, I. (1977). Belief, attitude, intention, and behavior: An introduction to theory and research. *Philosophy and Rhetoric*, *10*(2).
[15] Fishbein, M., & Ajzen, I. (2010). *Changing behavior: Theoretical considerations. Predicting and changing behavior: The reasoned action approach*. New York: Taylor & Francis Group.
[16] Halady, I. R., & Rao, P. H. (2010). Does awareness of climate change lead to behavioral change? *International Journal of Climate Change Strategies and Management*.
[17] Hernández-Perlines, F., Moreno-García, J., & Yáñez-Araque, B. (2017). Family firm performance: The influence of entrepreneurial orientation and absorptive capacity. *Psychology & Marketing*, *34*(11), 1057–1068.
[18] Hockerts, K., & Wüstenhagen, R. (2010). Greening Goliaths versus emerging Davids—Theorizing about the role of incumbents and new entrants in sustainable entrepreneurship. *Journal of Business Venturing*, *25*(5), 481–492.
[19] Kaesehage, K., Leyshon, M., Ferns, G., & Leyshon, C. (2019). Seriously personal: The reasons that motivate entrepreneurs to address climate change. *Journal of Business Ethics*, *157*(4), 1091–1109.
[20] Keogh, P. D., & Polonsky, M. J. (1998). Environmental commitment: A basis for environmental entrepreneurship? *Journal of Organizational Change Management*.
[21] Kolvereid, L. (1996). Prediction of employment status choice intentions. *Entrepreneurship Theory and Practice*, *21*(1), 47–58.
[22] Krueger Jr, N. F., Reilly, M. D., & Carsrud, A. L. (2000). Competing models of entrepreneurial intentions. *Journal of Business Venturing*, *15*(5–6), 411–432.
[23] Kuckertz, A., Berger, E. S., & Gaudig, A. (2019). Responding to the greatest challenges? Value creation in ecological startups. *Journal of Cleaner Production*, *230*, 1138–1147.
[24] Kuckertz, A., & Wagner, M. (2010). The influence of sustainability orientation on entrepreneurial intentions—Investigating the role of business

experience. *Journal of Business Venturing*, *25*(5), 524–539.
[25] Lans, T., Blok, V., & Wesselink, R. (2014). Learning apart and together: Towards an integrated competence framework for sustainable entrepreneurship in higher education. *Journal of Cleaner Production*, *62*, 37–47.
[26] Leiserowitz, A. A., Kates, R. W., & Parris, T. M. (2006). Sustainability values, attitudes, and behaviors: A review of multinational and global trends. *Annu. Rev. Environ. Resour.*, *31*, 413–444.
[27] Liñán, F., Rodríguez-Cohard, J. C., & Rueda-Cantuche, J. M. (2011). Factors affecting entrepreneurial intention levels: A role for education. *International Entrepreneurship and Management Journal*, *7*(2), 195–218.
[28] Loomes, G. (2006). (How) Can we value health, safety and the environment? *Journal of Economic Psychology*, *27*(6), 713–736.
[29] Lorenzoni, I., Nicholson-Cole, S., & Whitmarsh, L. (2007). Barriers perceived to engaging with climate change among the UK public and their policy implications. *Global Environmental Change*, *17*(3–4), 445–459.
[30] Lortie, J., & Castogiovanni, G. (2015). The theory of planned behavior in entrepreneurship research: What we know and future directions. *International Entrepreneurship and Management Journal*, *11*(4), 935–957.
[31] Maaloui, A., Perez, C., Bertrand, G., Razgallah, M., & Germon, R. (2018). "Cruel intention" or "entrepreneurial intention": What did you expect? An overview of research on entrepreneurial intention—An interactive perspective. *A Research Agenda for Entrepreneurial Cognition and Intention*, 7–46.
[32] Maibach, E., Roser-Renouf, C., & Leiserowitz, A. (2009). *Global warming's six Americas 2009: An audience segmentation analysis*.
[33] Mbebeb, F. E. (2012). Building ecological entrepreneurship: Creating environmental solutions based on the cultural realities and needs of local people. *J. Environ. Investig.*, *2*, 43–62.
[34] McMullen, J. S., & Shepherd, D. A. (2006). Entrepreneurial action and the role of uncertainty in the theory of the entrepreneur. *Academy of Management Review*, *31*(1), 132–152.
[35] Morales, C., Holtschlag, C., Masuda, A. D., & Marquina, P. (2019). In which cultural contexts do individual values explain entrepreneurship? An integrative values framework using Schwartz's theories. *International Small Business Journal*, *37*(3), 241–267.
[36] Moser, S. C. (2010). Communicating climate change: History, challenges, process and future directions. *Wiley Interdisciplinary Reviews: Climate Change*, *1*(1), 31–53.
[37] Muñoz, P., & Cohen, B. (2018). Sustainable entrepreneurship research: Taking stock and looking ahead. *Business Strategy and the Environment*, *27*(3), 300–322.
[38] Nisbet, M. C. (2009). Communicating climate change: Why frames matter for public engagement. *Environment: Science and Policy for Sustainable Development*, *51*(2), 12–23.
[39] Noseleit, F. (2010). The entrepreneurial culture: Guiding principles of the self-employed. In *Entrepreneurship and culture* (pp. 41–54). Springer.
[40] Nuringsih, K., Nuryasman, M. N., & IwanPrasodjo, R. A. (2019). Sustainable entrepreneurial intention: The perceived of triple bottom line among female students. *Jurnal Manajemen*, *23*(2), 168–190. https://doi.org/10.24912/jm.v23i2.472
[41] Nuringsih, K., & Puspitowati, I. (2017). Determinants of eco entrepreneurial intention among students: Study in the entrepreneurial education practices. *Advanced Science Letters*, *23*(8), 7281-7284. https://doi.org/10.1166/asl.2017.9351
[42] Parrish, B. D. (2010). Sustainability-driven entrepreneurship: Principles of organization design. *Journal of Business Venturing*, *25*(5), 510–523.
[43] Patchen, M. (2006). Public attitudes and behavior about climate change. *Purdue Climate Change Research Center Outreach Publication*, *601*.
[44] Peng, H., Li, B., Zhou, C., & Sadowski, B. M. (2021). How does the appeal of environmental values influence sustainable entrepreneurial intention? *International Journal of Environmental Research and Public Health*, *18*(3), 1070.
[45] Raducu, R., Soare, C., Chichirez, C.-M., & Purcarea, M. R. (2020). Climate change and social campaigns. *Journal of Medicine and Life*, *13*(4), 454.
[46] Riebl, S. K., Estabrooks, P. A., Dunsmore, J. C., Savla, J., Frisard, M. I., Dietrich, A. M., Peng, Y., Zhang, X., & Davy, B. M. (2015). A systematic literature review and meta-analysis: The Theory of Planned Behavior's application to understand and predict nutrition-related behaviors in youth. *Eating Behaviors*, *18*, 160–178.
[47] Saifi, B., & Drake, L. (2008). A coevolutionary model for promoting agricultural sustainability. *Ecological Economics*, *65*(1), 24–34.
[48] Schwartz, S. H. (1977). Normative influences on altruism. In *Advances in experimental social psychology* (Vol. 10, pp. 221–279). Elsevier.
[49] Schwartz, S. H. (2011). *Values: Cultural and individual*. In F. J. R. van de Vijver, A. Chasiotis, & S. M. Breugelmans (Eds.), *Fundamental questions in cross-cultural psychology* (pp. 463–493). Cambridge University Press.
[50] Schwartz, S. H., Cieciuch, J., Vecchione, M., Davidov, E., Fischer, R., Beierlein, C., Ramos, A., Verkasalo, M., Lönnqvist, J.-E., & Demirutku, K. (2012a). Refining the theory of basic individual values. *Journal of Personality and Social Psychology*, *103*(4), 663.
[51] Schwartz, S. H., Cieciuch, J., Vecchione, M., Davidov, E., Fischer, R., Beierlein, C., Ramos, A., Verkasalo, M., Lönnqvist, J.-E., & Demirutku, K. (2012b). Refining the theory of basic individual values. *Journal of Personality and Social Psychology*, *103*(4), 663. https://doi.org/10.1037/a0029393
[52] Shaikh, S. L., Sun, L., & van Kooten, G. C. (2007). Treating respondent uncertainty in contingent valuation: A comparison of empirical treatments. *Ecological Economics*, *62*(1), 115–125.
[53] Shane, S., & Venkataraman, S. (2000). The promise of entrepreneurship as a field of research. *Academy of Management Review*, *25*(1), 217–226.
[54] Shepherd, D. A., Kuskova, V., & Patzelt, H. (2009). Measuring the values that underlie sustainable development: The development of a valid scale. *Journal of Economic Psychology*, *30*(2), 246–256.
[55] Shepherd, D. A., & Patzelt, H. (2011). The new field of sustainable entrepreneurship: Studying entrepreneurial action linking "what is to be sustained" with "what is to be developed." *Entrepreneurship Theory and Practice*, *35*(1), 137–163.
[56] Sher, A., Abbas, A., Mazhar, S., Azadi, H., & Lin, G. (2020). Fostering sustainable ventures: Drivers of sustainable start-up intentions among aspiring entrepreneurs in Pakistan. *Journal of Cleaner Production*, *262*, 121269.
[57] Shi, Y., Yuan, T., Bell, R., & Wang, J. (2020). Investigating the relationship between creativity and entrepreneurial intention: The moderating role of creativity in the theory of planned behavior. *Frontiers in Psychology*, *11*, 1209.
[58] Spash, C. L. (2002). Informing and forming preferences in environmental valuation: Coral reef biodiversity. *Journal of Economic Psychology*, *23*(5), 665–687.
[59] Stubbs, W. (2017). Sustainable entrepreneurship and B corps. *Business Strategy and the Environment*, *26*(3), 331–344.
[60] Tan, C. K., Ogawa, A., & Matsumura, T. (2008). Innovative climate change communication: Team minus 6%. *Global Environment Information Centre (GEIC), United Nations University (UNU)*, 53–70.
[61] Thelken, H. N., & de Jong, G. (2020). The impact of values and future orientation on intention formation within sustainable entrepreneurship. *Journal of Cleaner Production*, *266*, 122052.
[62] Thøgersen, J., & Ölander, F. (2002). Human values and the emergence of a sustainable consumption pattern: A panel study. *Journal of Economic Psychology*, *23*(5), 605–630.
[63] Tilley, F., & Young, W. (2009). Sustainability Entrepreneurs. *Greener Management International*, *55*.
[64] Timm, S. N., & Deal, B. M. (2016). Effective or ephemeral? The role of energy information dashboards in changing occupant energy behaviors. *Energy*

*Research & Social Science*, *19*, 11–20.
[65] Van Oorschot, J. A., Hofman, E., & Halman, J. I. (2018). A bibliometric review of the innovation adoption literature. *Technological Forecasting and Social Change*, *134*, 1–21.
[66] Yuriev, A., Dahmen, M., Paillé, P., Boiral, O., & Guillaumie, L. (2020). Pro-environmental behaviors through the lens of the theory of planned behavior: A scoping review. *Resources, Conservation and Recycling*, *155*, 104660.
[67] Zhao, H., Seibert, S. E., & Lumpkin, G. T. (2010). The relationship of personality to entrepreneurial intentions and performance: A meta-analytic review. *Journal of Management*, *36*(2), 381–404.